\documentclass[twocolumn,showpacs,preprintnumbers,amsmath,amssymb]{revtex4}

\usepackage{graphicx}
\usepackage{dcolumn}
\usepackage{bm}

\begin{document}


\title{Topological Order in Frustrated Josephson Junction Arrays}
\author{M. Cristina Diamantini}
\email{cristina.diamantini@pg.infn.it}
\affiliation{%
INFN and Dipartimento di Fisica, University of Perugia, via A. Pascoli, I-06100 Perugia, Italy
}%

\author{Pasquale Sodano}
\email{pasquale.sodano@pg.infn.it}
\affiliation{%
INFN and Dipartimento di Fisica, University of Perugia, via A. Pascoli, I-06100 Perugia, Italy
}%

\author{Carlo A. Trugenberger}
\email{ca.trugenberger@bluewin.ch}
\affiliation{%
SwissScientific.com, ch. Diodati 10 CH-1223 Cologny, Switzerland
}%


\date{\today}

\begin{abstract}
We show that electrically and magnetically frustrated Josephson junction arrays
(JJAs) realize topological order with a non-trivial ground state degeneracy on
manifolds with non-trivial topology. The low-energy theory has the
same gauge dynamics of the unfrustrated JJAs but for different, "fractional" 
degrees of freedom, a principle
reminescent of Jain's composite electrons in the fractional quantum Hall effect.

\end{abstract}
\pacs{11.10.z;74.20.Mn;05.30.Pr}

\maketitle
The idea of quantum order has its origin in the discovery of the fractional quantum Hall
liquids \cite{Lau}, that are incompressible and exist only at some
"magical" filling fractions. The internal
order characterizing these states is a new type of order, called topological order,
different from any other previously known type of order and it is not based on spontaneous symmetry
breaking.
Quantum order describes in general the zero-temperature properties of a quantum
ground state and characterizes universality classes of quantum
states, described by {\it complex} ground state wave-functions
\cite{wen}.
Topological order is a special type of quantum order
whose hallmarks are the presence of a gap for all excitations
(incompressibility) and the degeneracy of the ground state on
manifolds with non-trivial topology \cite{wen}. In the case of the
fractional quantum Hall effect, which is a (2+1)-dimensional system,
another hallmark is the presence of excitations with fractional spin
and statistics, called anyons \cite{wilczek}. The long-distance
properties of these topological fluids can be explained by an
infinite-dimensional $W_{1+\infty}$ dynamical symmetry \cite{czt},
and are described by effective Chern-Simons field theories
 \cite{wen2} with compact gauge group, which break P-and
T-invariance.
In fact, the simplest example of a topological fluid \cite{fwz} is a ground-state
described by a low-energy effective action given solely by the
topological Chern-Simons term \cite{jackiw} $S = k/4\pi\ \int
d^3x \ A_{\mu} \epsilon^{\mu \nu \alpha}
\partial_{\nu} A_{\alpha}$ for a compact $U(1)$ gauge field
$A_{\mu}$ whose dual field strength $F^{\mu} = \epsilon^{\mu \nu
\alpha} \partial_{\nu} A_{\alpha}$ is interpreted as the conserved matter
current. In this case the degeneracy of the ground state on a
manifold of genus $g$ will be $k^g$ (or $(k_1 k_2)^g$ if $k =
k_1/k_2$ is a  rational number). Topological order in conventional
superconductors has been investigated in \cite{hansson}.

In \cite{nsm} we proposed a new superconductivity mechanism
which is based on a topologically ordered
ground state rather than on the usual Landau mechanism of
spontaneous symmetry breaking. Contrary to anyon superconductivity
it works in any dimension and it preserves P-and T-invariance. 
The topologically ordered superconductors we proposed have a
long-distance hydrodynamic action which can be entirely formulated
in terms of generalized compact gauge fields, the dominant term
being the topological BF action \cite{birmi}.  
In (2+1) dimensions, it reduces to a mixed Chern-Simons model
\cite{freedman}. 

A first concrete example of topological superconductors is provided by 
JJAs.
In \cite{dst}, we have proven that planar JJAs can be exactly 
mapped onto an Abelian gauge theory with a
mixed Chern-Simons term (BF-model) which complete determines the low energy
dynamics.  The Abelian gauge
theory exactly reproduces the phase diagram of JJAs and the
insulator/superconductor quantum phase transition at $T=0$
\cite{jja}. JJAs, {\bf and fully frustrated JJA ladders \cite {napo}}, have also been recently  considered by several
other authors \cite{as, joffe}  as  devices 
exhibiting topological order.
For planar unfrustrated JJAs, however, the topological fluid is described by a $k=1$ mixed
Chern-Simons model and, thus, there is
no degeneracy of the ground state \cite{dst,nsm}.

In \cite{bff} we argued that frustrated JJAs may support a
topologically ordered ground state with non-trivial degeneracy on
higher genus surfaces.
In this paper we give an exact derivation of the gauge theory that
describes Josephson junction arrays in presence of electric and magnetic
frustration. We will then show, for rational frustration, how this can be
rewritten as the same dynamics of the unfrustrated model but for different
degrees of freedom, a principle analogous to Jain's composite electrons
familiar from the fractional quantum Hall effect. The only difference is the
Chern-Simons coefficient, determined by the denominator of the frustration: as we
anticipated  this leads to a ground state degeneracy for arrays with non-trivial
topology.
 
We shall consider JJAs fabricated on a square planar lattice of
spacing $l = 1$ made of superconducting islands with nearest neighbours
Josephson couplings of strength $E_J$ \cite{jja}. Each island has a
capacitance $C_0$ to the ground; moreover, there are nearest
neighbours capacitances $C$. To implement a torus topology we impose
doubly periodic conditions at the boundary of the square lattice.
The Hamiltonian describeing the system is, in the limit $C\gg C_0$:
\begin{equation}H= \sum_{\bf x} \ N^2E_C \ p_0 {1\over -\Delta } p_0 +
\sum _{{\bf x},i} \ E_J \left( 1-{\rm cos} \ N\left( \Delta _i \Phi \right)
\right) \ ,
\label{hjjc}
\end{equation}
where $E_C\equiv q_e^2/2C$. 
The phases $\Phi$ are quantum-mechanically conjugated to the charges
$Q$ on the islands: these are quantized in integer multiples of $N$
(Cooper pairs for $N=2$): $Q= q_e N p_0$, $p_0 \in Z$
where $q_e$ is the electron charge. 
In \cite{dst} we have shown that the zero-temperature partition
function of JJAs may be written in terms of two effective gauge
fields $A_{\mu}$ (vector) and $B_{\mu}$ (pseudovector). In the low energy limit the partition
function is
\begin{eqnarray}
Z &&= \sum_{\{ Q_0 \} \atop \{ M_0 \} } \left( \delta M_0 \right)\
\int {\cal D}A_{\mu } \int {\cal D}B_{\mu } \ {\rm exp}(-S)\ ,
\nonumber \\
S &&= \int dt \sum_{\bf x} -i{ 1\over 2\pi }\ A_{\mu }K_{\mu \nu }B_{\nu } +\nonumber \\
&&+i A_0 Q_0 + i  B_0 M_0\ .
\label{ac}
\end{eqnarray}
This form of the partition function holds true also with toroidal
boundary conditions. The gauge fields, with compactification radius $2 \pi$, embody the original degrees of
freedom through their dual field strengths $p_{\mu} \propto K_{\mu \nu}
B_{\nu}$ and $q_{\mu} \propto K_{\mu \nu} A_{\nu}$ representing,
respectively, the conserved charge (vector) current and the
conserved vortex (axial) current. $K_{\mu \nu}$ is the
 lattice Chern-Simons term
\cite{latticecs}, defined by $K_{00} = 0$, $K_{0i} = -\epsilon_{ij}
d_j$, $K_{i0} = S_i \epsilon_{ij} d_j$ and $K_{ij} = -S_i
\epsilon_{ij} \partial_0$, in terms of forward (backward) shift and
difference operators $S_i$ ($\hat S_i$) and $d_i$ ($\hat d_i$). Its
conjugate $\hat K_{\mu \nu}$ is defined by $\hat K_{00} = 0$, $\hat
K_{0i} = - \hat S_i \epsilon_{ij} \hat d_j$, $\hat K_{i0} =
\epsilon_{ij} \hat d_j$ and $\hat K_{ij} = - \hat S_j \epsilon_{ij}
\partial_0$. The two Chern-Simons kernels $K_{\mu \nu}$ and $\hat
K_{\mu \nu}$ are interchanged upon integration (summation) by parts
on the lattice.

The topological excitations are described by the integer-valued
fields $Q_0$ and $M_0$ and represent unit charges and vortices
rendering the gauge field components $A_0$ and $B_0$ integers via
the Poisson summation formula; their fluctuations determine the
phase diagram \cite{dst}. In the classical limit
the magnetic excitations are dilute and the charge excitations condense,
rendering the system a superconductor: vortex confinement amounts
here to the Meissner effect. In the quantum limit, the magnetic
excitations condense while the charged ones become dilute: the
system exhibits insulating behavior due to vortex superconductivity
accompanied by a charge Meissner effect.

In presence of an homogeneous charge  frustration, represented by external
charges $p_f$ on the islands (\ref{hjjc}) becomes:
\begin{eqnarray}H= &&\sum_{\bf x} \ N^2E_C \ \left(p_0 - p_f \right) 
{1\over -\Delta } \left(p_0 - p_f \right)  +\nonumber \\
&&\sum _{{\bf x},i} \ E_J \left( 1-{\rm cos} \ N\left( \Delta _i \Phi \right)
\right) \ ,
\label{efhjjc}
\end{eqnarray}
Following the same steps as in \cite{dst} we can write the partition function for the
JJAs as:
\begin{eqnarray}&&Z = \sum _{\{ Q_{0 } \}} \int {\cal D}a_{\mu }
\int {\cal D}b_{\mu } \ {\rm exp}(-S) \ ,\nonumber \\
&&S = \int d t \sum_x -i2\pi \ a_{\mu }K_{\mu \nu }b_{\nu } + 
{p_i^2\over 2l_0E_J} + \nonumber \\
&&+ N^2l_0E_C \ \left(p_0 - p_f \right) {1\over
-\Delta }\left(p_0 - p_f \right) + i2\pi a_{0 }Q_{0 } \ ,
\label{pfjje}
\end{eqnarray}
with $p_{\mu } \equiv K_{\mu \nu }b_{\nu } \ , b_{\mu }\in R\ ,
q_{\mu } \equiv \hat K_{\mu \nu } a_{\nu }\ , a_{\mu }\in Z$ 

In this representation $K_{\mu \nu }b_{\nu }$ represents the conserved
three-current of charges, while $\hat K_{\mu \nu }a_{\nu }$ represents
the conserved three-current of vortices. $a_\mu$ and $b_\mu$ are compact gauge
fields with period 1. The third term in the action
(\ref{pfjje})\ contains two parts: the longitudinal
part $\left( p_i^L \right) ^2$ describes the
Josephson currents and represents a kinetic term for the charges;
the transverse part $\left( p_i^T \right) ^2$ can be rewritten as
a Coulomb interaction term for the vortex density $q_0$ by solving
the Gauss law enforced by the Lagrange multiplier $b_0$.

The partition function (\ref{pfjje})\ displays
a high degree of symmetry between the charge and the vortex
degrees of freedom.
The only term which breaks this symmetry (apart from the integers $Q_{0 }$)
is encoded in the kinetic term for the charges (Josephson currents).
Following \cite{dst} we introduce the
self-dual approximation of JJAs by adding to the action in (\ref{pfjje})
a bare kinetic term for the vortices
$\sum_x {\pi ^2 \over N^2 l_0
E_C} q_i^2$. The coefficient is chosen so that the transverse
part of this term reproduces exactly the Coulomb term for the charges
upon solving the Gauss law enforced by the Lagrange multiplier $a_0$.

In order to exactly reproduce the Coulomb term, including the frustration, we need to
add an extra $i2\pi a_0p_f $.
The action then becomes:
\begin{eqnarray}S_{SD} = \int d t &&\sum_x -i2\pi \ a_{\mu }K_{\mu \nu }b_{\nu } + {p_i^2\over 2l_0E_J}
+{\pi ^2 q_i^2 \over N^2l_0E_C} +\nonumber \\
 &&i2\pi a_0p_f + i2\pi a_{0}Q_{0 } +i 2\pi b_{0}
M_{0 } \ ,
\label{efsd}
\end{eqnarray}
where we are also forced to introduce new integers $M_{0 }$
via the Poisson formula to guarantee that the charge current $K_{\mu \nu }
b_{\nu }$ remains an integer (note that, without the kinetic term for the
vortices, both 
conserved currents are integers: indeed, the
summation over $\{ Q_{0} \}$ makes $a_{0 }$ , and then
the summation over $\{ a_{0 }
\}$ makes $K_{\mu \nu}b_{\nu }$ an integer).
Since the properties of the system are periodic with respect to the charges, we
can restrict $-1/2 \leq  p_f \leq 1/2$ \cite{otto}:  henceforth we have
 $p_f = p/q$.
Since we are intersted only on the ground state, we will consider only to the
low energy limit. 
In this limit the partition function becomes:
\begin{eqnarray}
Z_{LE} &&= \sum_{\{ Q_0 \}\atop \{ M_0 \} } \delta \left(M_0 \right) \
\int {\cal D}a_{\mu } \int {\cal D}b_{\mu } \ {\rm exp}\ (-S_{LE})\ ,
\nonumber \\
S_{LE} &&= \int dt \sum_{\bf x} -i2\pi \ a_{\mu }K_{\mu \nu }b_{\nu }\nonumber \\
&&+ i2\pi a_0 Q_0 + i2\pi b_0 M_0 + i2\pi a_0 {p \over q}\ .
\label{af}
\end{eqnarray}

The effect of charge frustration, the last term in (\ref{af}), can be reabsorbed
as follows. We first change the compactification radius of the gauge field
$a_0$ from the standard 1 to $q$. As usual, this corresponds to admitting
fractional charges $N/q$ in the model (in order to keep with standard notation we
rescale $a_\mu \rightarrow q a^{'}_\mu \ ,$ where $a_\mu$ has again period 1).
Correspondingly we have to rescale the magnetic topological excitations $M_0
\rightarrow q M_0$ in order to be able to absorb integer shifts in the rescaled
$a_\mu$. This corresponds to having a condensate of fractionally
charged particles $N/q$ while the magnetic excitations have fractional (in units
of the magnetic flux quantum  $2\pi q/N$) vortices $2\pi/N$. Since the magnetic
topological excitations $q M_0$ represents tunneling between vortex sectors 
differing by $q$ units $2\pi/N$ we have a $Z_q$ theory in which fractional
vortices and charges have mutual fractional statistics $2\pi/q$ \cite{chia}. For
$q=2$ such mutual anyons where considered by Kitaev \cite{kit} in his model of
topological quantum computation. At this point the remaning frustration effect
ca  be easily accounted by a redefinition $$b_0^{'} = b_0;  b_i^{'} = 
b_i + A_i^{\rm ext} \ ,$$ where we have written, without loss of generality, the
frustration term $p_f$ as $ p_f = K_{0i}A_i^{\rm ext}$. 
The action (\ref{af}) becomes
\begin{eqnarray}S_{LE} &&= \int dt \sum_{\bf x} -i2\pi q\ 
a_{\mu }K_{\mu \nu }b_{\nu }\nonumber \\
&&+ i2\pi q a_0 Q_0 + i2\pi q b_0 M_0 + i2\pi q a_0 K_{0i}A_i^{\rm ext}\ .
\label{afs}
\end{eqnarray}
Since $A_i^{\rm ext}$ is time-independent we have $a_iK_{ij }b_j^{'} \rightarrow 
a_iK_{ij }b_j$ and we get
\begin{eqnarray}
Z_{LE} &&= \sum_{\{ Q_0 \}\atop  \{M_0\}} \ \delta \left( M_0 \right) \
\int {\cal D}a_{\mu } \int {\cal D}b_{\mu } \ {\rm exp}\ (-S_{LE})\ ,
\nonumber \\
S_{LE} &&= -i \int dt \sum_{\bf x} 2\pi q a_0 K_{0i}  b_i^{'} + 2 \pi q a_i K_{i0} b_0
+\nonumber \\
&& 2 \pi  q a_i K_{ij} b_j^{'} + i2\pi q b_0 M_0 + i2  \pi q a_0 Q_0 \ . \label{ai}
\end{eqnarray}

Finally, by rescaling $A_\mu = 2 \pi a_\mu,\ B_\mu = 2 \pi b_\mu^{'}$ 
in (\ref{ai}), we obtain the low energy partition function:
\begin{eqnarray}
Z_{LE} &&= \sum_{\{ Q_0 \}\atop  \{M_0\}} \ \delta \left( M_0 \right) \
\int {\cal D}A_{\mu } \int {\cal D}B_{\mu } \ {\rm exp}\ (-S_{LE})\ ,
\nonumber \\
S_{LE} &&= -i\int dt\sum_{\bf x} {q \over 2 \pi}A_\mu K_{\mu \nu}  B_\nu +
\nonumber \\
&&i q B_0 M_0 + i q A_0 Q_0 \ . \label{aaim}
\end{eqnarray}
This is the same action as for the unfrustrated case (same periodicity of the
gauge fields by construction) with the only difference of a Chern-Simons
coefficient $q$ determined by the denominator of the frustration. In analogy to
the Jain construction familiar from the QHE we have the same dynamics but for different
degrees of freedom. This, however,  affects the ground state degeneracy that is
given \cite{gord} by $(k)^g = q$ on the torus.

The magnetic frustration can be treated in a similar way. In presence of a
uniform magnetic frustration the action \ref{hjjc} becomes:
\begin{eqnarray}H= &&\sum_{\bf x} \ N^2E_C \ p_0  {1\over -\Delta } p_0   +\nonumber \\
&&\sum _{{\bf x},i} \ E_J \left( 1-{\rm cos} \ \left( N \Delta _i \Phi - 
A_i^{\rm ext}\right) \right) \ ,
\label{mfhjjc}
\end{eqnarray}
with $A_i^{\rm ext}$ time independent and such that $\sum_{\rm plaquette}
A_i^{\rm ext} = 2\pi f$ where $2\pi f$ is the flux piercing the elementary
plaquette.
Due to the periodicity properties of the system \cite{otto}, we can restrict
$0 \leq  p_f \leq 1/2$ and  rewrite: $f = p/q$.
Following again the same steps as in \cite{dst}, we arrive at
\begin{eqnarray}Z &&= \sum _{\{ v_{\mu } \} } \int
{\cal D}p_{\mu } \int {\cal D}\Phi \ {\rm exp}(-S)\ ,\nonumber
\\
S &&= \int d t \sum_x -iN p_{\mu } \left( \Delta _{\mu } \Phi + {2\pi \over N} v_{\mu }
\right) + i p_i A_i^{\rm ext} + \nonumber \\
&&+  iN^2 E_C l_0 \ p_0 {1\over -\Delta} p_0 +
{p_i^2\over 2l_0E_J} \ .
\label{mpfjjb}
\end{eqnarray}
Introducing the compact gauge fields $a_\mu$ and $b_\mu$, both with period 1, such that
$K_{\mu \nu }b_{\nu }$ represents the conserved
three-current of charges,and $\hat K_{\mu \nu }a_{\nu }$ represents
the conserved three-current of vortices, we can rewrite the action in 
(\ref{mpfjjb}) as
\begin{eqnarray}S &&= \int d t\sum_x -i2\pi \ a_{\mu }K_{\mu \nu }b_{\nu } + 
{\pi ^2 q_i^2 \over N^2l_0E_C}+ {p_i^2\over 2l_0E_J} + \nonumber \\
&&+ i b_0 \hat K_{0 i } A_i^{\rm ext} +
i2\pi a_{0 }Q_{0 } + i2\pi b_{0 }M_{0 }  \ ,
\label{mpfjjd}
\end{eqnarray}
where we have introduced the two integer fields $M_0$ and $Q_0$ as in (\ref{ac}). 

The low energy partition function is:
\begin{eqnarray}
Z_{LE} &&= \sum_{\{ Q_0 \}\atop \{ M_0 \} } \delta \left(M_0 \right) \
\int {\cal D}a_{\mu } \int {\cal D}b_{\mu } \ {\rm exp}\ (-S_{LE})\ ,
\nonumber \\
S_{LE} &&= \int dt \sum_{\bf x} -i2\pi \ a_{\mu }K_{\mu \nu }b_{\nu }\nonumber \\
&&+ i2\pi a_0 Q_0 + i2\pi b_0 M_0 + i2\pi b_0 {p \over q}\ .
\label{afm}
\end{eqnarray}

Following the same steps as for charge frustration we reabsorb  
the effect of magnetic frustration, the last term in (\ref{afm}),
as follows. We first change the compactification radius of the gauge field
$b_0$ from the standard 1 to $q$; this corresponds to admitting
fractional fluxes $2\pi/Nq$.We then
rescale $b_\mu \rightarrow q b^{'}_\mu \ ,$ where $b_\mu$ has again period 1.
Correspondingly, we have to rescale the charge topological excitations $Q_0
\rightarrow q Q_0$ in order to be able to absorb integer shifts in the rescaled
$b_\mu$. In this case we have  fractional
magnetic fluxes $2\pi/Nq$ while the charge excitations have fractional (in units
of the electric charge quantum  $N q$) charge $N$. Since the charge
topological excitations $q Q_0$ represents tunneling between charge sectors 
differing by $q$ units $N$ we have a $Z_q$ theory, in agreement with the result
found in \cite{ds}, in which fractional
vortices and charges have mutual fractional statistics $2\pi/q$. 
We now change variables: 
$$a_0^{'} = a_0 ;  a_i^{'} = a_i + A_i^{\rm ext}\ ,$$
with $A_i^{\rm ext}$  time-independent and rewrite the action 
in (\ref{afm}) as:
\begin{eqnarray}
S_{LE} &&= \int dt \sum_{\bf x} -i2\pi q \ a_{\mu }^{'}K_{\mu \nu }b_{\nu }
+ i2 q \pi a_0^{'} Q_0 \nonumber \\
&&+ i2 q \pi b_0 M_0   \ .
\label{mafs}
\end{eqnarray}

By rescaling $A_\mu = 2 \pi a_\mu^{'},\ B_\mu = 2 \pi b_\mu$ in (\ref{mafs})
we obtain the low energy partition function:
\begin{eqnarray}
&&Z_{LE} = \sum_{\{ Q_0 \}\atop  \{M_0\}} \ \delta \left( M_0 \right) \
\int {\cal D}A_{\mu } \int {\cal D}B_{\mu } \ {\rm exp}\ (-S_{LE})\ ,
\nonumber \\
&&S_{LE} = - i \int dt \sum_{\bf x} {q \over 2 \pi}A_\mu K_{\mu \nu}  B_\nu + 
i q B_0 M_0 + i q A_0 Q_0 \ . \label{aaimm}
\end{eqnarray}
Again this is the same action as for the unfrustrated case (same periodicity of the
gauge fields by construction). The only change is again the Chern-Simons
coefficient $q$ determined by the denominator of the frustration,
giving a ground state degeneracy $(k)^g = q$ on the torus \cite{gord}.

{\bf Planar frustrated Josephson junction arrays with toroidal boundary
conditions represent  the simplest example of
a superconductor that exibits topological order with  non-trivial ground 
state degeneracies.
Furthermore they are accessible experimentally \cite{cirillo} and thus they may be used to test the
topological superconductivity mechanism proposed in \cite{bff}.}

The result we have obtained is very similar to the one obtained in \cite{wwz} 
for the the frustrated Heisenberg model. Here the frustration is obtained by introducing next to
nearest-neighbors interactions. For the chiral spin phase, the low energy limit is described by
a Chern-Simons theory with coefficient $k = q$ if the effective flux threading the plaquette
is $\phi =  2 \pi {p \over q}$. This is true also for $\phi = \pi$, because due to  
the frustration there is an effective flux $\pi/2$ on triangular plaquettes.
This hints to a possible equivalence between geometric frustration and frustration
induced by an external field

\end{document}